# Millimeter Wave Dual-Band Multi-Beam Waveguide Lens-Based Antenna

Vedaprabhu Basavarajappa[1, 2], Alberto Pellon[1], Ana Ruiz[1], Beatriz Bedia Exposito[1]

Lorena Cabria[1] and Jose Basterrechea[2]

[1]Dept. of Antennas, TTI Norte
Parque Científico y Tecnológico de Cantabria,
C/ Albert Einstein nº 14, 39011
Santander, Spain

[2]Departamento de Ingeniería de Comunicaciones,
Universidad de Cantabria
Edificio Ingeniería de Telecomunicación Profesor José Luis García García
Plaza de la Ciencia, Avda de Los Castros, 39005, Santander, Spain

Email: {veda, apellon, aruiz, bbedia, lcabria} @ttinorte.es, jose.basterrechea@unican.es

*Abstract*— **A multi-beam antenna with a dual band operation in the 28 GHz and 31 GHz millimeter wave band is presented. The antenna has a gain of around 15 dBi in each of the three ports. The spatial footprint of the antenna is 166 mm x 123 mm x 34 mm. A waveguide lens-based approach is used to attain this gain. Cylindrical to planar wavefront transformation by a phase extraction and compensation method drives the design of the antenna. The dual band operation of the antenna aids in transmitting and receiving at two independent frequencies. Three beams originating from a shared aperture are designed to target directions of -60°, 0° and 60°. These features make the antenna a potential candidate for 5G millimeter wave applications.**

## I. Introduction

In applications requiring a wide-angle scanning, wide band or dual-band coverage such as the deployment of antennas in a 5G millimeter wave small cell scenario of Local Multipoint Distribution System (LMDS) or in a constellation-based satellite system, there is a requirement for antennas that can cast multiple beams with a high directive gain. The high gain in the antenna beam can primarily be achieved either through a massive array of antennas or through a lens based focusing scheme. The latter is appealing owing to the simplicity of its design and the ease with which the beam can be steered to different angles using the same lens. The lens antenna is generally made of dielectric or transmission line delays. In [1] a metallic delay lens is introduced in which the focusing action is obtained by reduction of the phase velocity of the radio waves passing through the lens. In comparison to [1] a method of increasing the phase velocity of the radio waves is proposed in metal plate lens such as the Ruze lens [2] and the Rotman lens [3]. An approach to distribute the radiating source elements along a focal surface of a planar EM lens is discussed in [4], which is termed as the lens antenna array. The angle dependent energy focusing property of lens antenna array is harnessed in improving the channel capacity of mmWave channel. Another approach in realizing delay lens is using a transmission line delay that emulates the various phases as along a lens [5]. Here the profile of the transmission line is chosen in a such a way that a line source placed inside the lens radiates a plane wave on the outside before being launched. A multi-layer leaky wave-based solution based on the substrate integrated waveguide (SIW) technology is proposed in [6]. The method consists of launching the source waves through a SIW and coupling it to the radiating leaky wave slots, with a quasi-optical system of EM coupling. [7] introduces a parallel plate waveguide-based solution to the multi-beam casting problem. A continuous delay lens based quasi optical beamformer was designed, manufactured and tested for operation in the Ku band. [8] provides the analysis of the design in [7] using a ray tracing technique. A detailed numerical analysis tool is provided in [9] to predict the performance of the beamformer.

The papers [1-6] introduce diverse ways of realizing the energy focusing property with SIW, transmission line and lens technology. [7-9] propose a way to design a beamformer- all in the sub-20 GHz frequency range using a ridge-based lens. The approach however in these methods is to use ray optics to arrive at the position of the feeding horns and to decide on the contour of the lens. In this paper, a proposition is made to design the feeding horn and ridge-based lens using a phase extraction and compensation method, building solely upon full wave simulations. The use of an initial simpler full wave simulation to extract information on the phases of the wavefront helps in the easier refining of the design model to attain beam focusing. In the approach in [9] an analytical formulation is made based on ray optics to predesign the beamformer, which is subject to local and global optimization procedures to fine tune the beamformer. In the proposed design the final optimization is minimized owing to the individual considerations and co-design of the parts that make up the beamformer. Another key difference is to use E plane sectoral horns that aid in the ease of integration to the parallel plate waveguide. The antenna has also been designed for dual band operation in the Ka band as opposed to [7-9] which gives it scope to act as transceiver for Tx and Rx at two different frequencies. Also, the designed uniform height of the ridge lens makes the 3D fabrication much simpler as compared to [7] since there is no tapering of the ridge towards the edges. The parallel plate waveguide beamformer is designed and performance validated by full-wave simulation for operation in the dual band of 28 GHz and 31 GHz, with a bandwidth of 1 GHz each. A systematic design approach encompassing the co-design of the sectoral E-plane feed horns, the parallel plate waveguide, the ridge lens and the flared horn aperture is presented. The multi-beam angular directions have been validated with directional gains around 15 dBi targeted in the directions of -60°, 0° and 60°.

The paper is organized as follows: Section I introduces the prior and related work and highlights the features of the presented work in comparison to them. Section II deals with the design principle of the antenna and explains the mechanism of the antenna operation. Section III focuses on the detailed antenna design. Section IV presents the simulation results of the three port antenna characteristics that include the return losses, port-to-port isolation, radiation efficiency and the gain pattern plots. Section V concludes, summarizing the results achieved and the potential applications of the antenna.

## II. DESIGN PRINCIPLE

The parallel plate waveguide lens-based antenna working principle is discussed in this section. The gain of a E plane sectoral horn antenna designed around the frequency of 28 GHz and 31 GHz is about 10.5 dBi. The lens-like operation performed on the $E_x$ field vector of the feeding horns along with the flared aperture results in a gain increase of about 6 dBi. Essentially, the ridge-based lens-like transformation of the cylindrical wavefront to the planar wavefront is the cause for the increase in the gain. This can be regarded as beam-focusing through transformation of wavefronts. The principle of operation with analogy drawn to a simple ray tracing technique is described in the following section. The transversal section of the waveguide lens antenna is shown in Fig. 1. The transformation of the spatial amplitude distribution and the dependent phase distribution from cylindrical to planar wavefront for the $E_x$ field within the waveguide lens is shown in Fig. 2 for the three ports of excitation. A gradual change of the wavefronts along the periphery of the ridge is seen. The planar wavefront thus generated is responsible for projecting the high gain beams aided by the flared aperture.

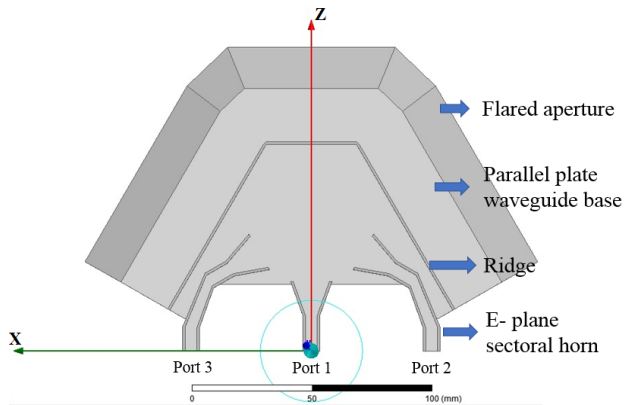

Fig. 1. Transversal section of the waveguide lens antenna

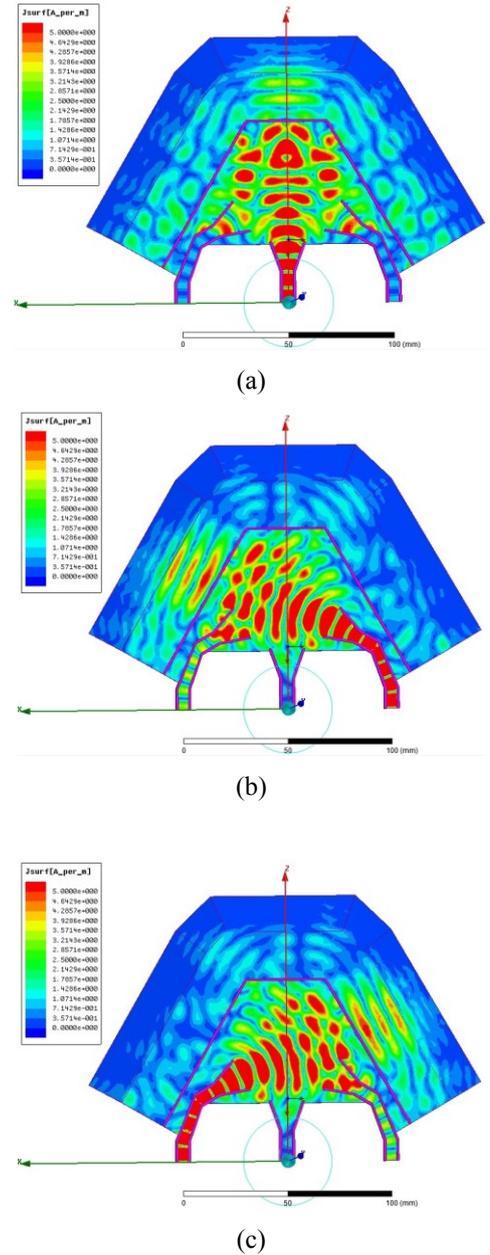

Fig. 2. Spatial distribution of the amplitude of surface current also indicating the wavefront shape evolution through the ridge for (a) Port 1 (b) Port 2 and (c) Port 3

## III. ANTENNA DESIGN

The antenna design consists of the co-design of a conglomeration of four interdependent design parts - namely the parallel plate waveguide, the feeding horns, the ridge-based lens and the flared radiating aperture. The following section describes the design procedure of each of these parts in separate and in unison. A 3D perspective view of the geometry of the antenna is shown in Fig. 3.

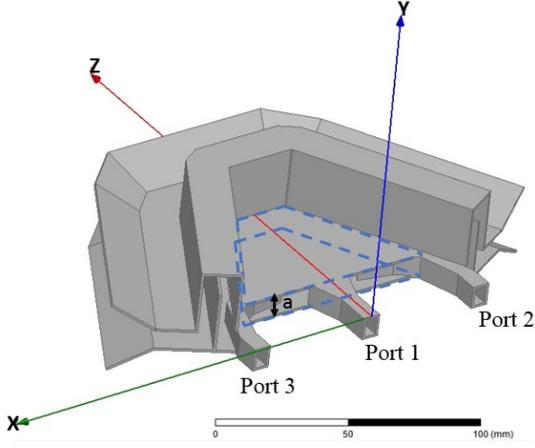

Fig. 3. 3D perspective view of the dual band multi-beam antenna

*A. Design of the parallel plate waveguide*

The parallel plate waveguide (PPW) structure is chosen to act as the shared guiding interface between the three launching ports and their corresponding wave propagation. Following the axis notation in Fig. 3, for the excitation of port 1, the parallel plate extension along the x direction can be theoretically considered to be infinite, which in the proposed scenario depicts resulting radiation to free space along the z direction. The separation between the plates 'a' is a critical factor in ensuring the propagation of the guided wave along the direction z. This is derived using the following steps to ensure propagation for frequencies above 20 GHz. The antenna as shown in Fig. 3 is placed with the parallel plate waveguide plates placed parallel to the xz plane, with the plate separation distance 'a' along the y axis. In the above scenario, considering excitation along port 1, the Maxwell´s wave equation becomes:

$$\nabla^2 \mathbf{E} + \omega^2 \mu\varepsilon \mathbf{E} = 0 \quad (1)$$

As stated earlier, the plates can be considered to be infinite in the x direction and it is necessary to consider the propagation in the z direction. Therefore, component $E_x$ only is assumed along with the additional conditions that the partial derivative along x is zero and along y and z is non-zero. These imposed assumptions define the TE modes and the differential wave equation is used to obtain the classical dispersion relation [10]:

$$\beta_z^2 + \beta_y^2 = \omega^2\mu\varepsilon = \beta^2, \text{ where } \beta_y = \frac{m\pi}{a} \quad (2)$$

where index m=1, 2... which on solving for $\beta_z$ yields

$$\beta_z = \sqrt{\omega^2\mu\varepsilon - \left(\frac{m\pi}{a}\right)^2} \quad (3)$$

The first TE mode is the $TE_1$ mode. Since the wave must propagate in the positive z direction, for the propagation to physically occur $\beta_z^2 > 0$; following this the guidance condition was deducted as:

$$f > \frac{m}{2a\sqrt{\mu\varepsilon}} \quad (4)$$

that gives the limit for the cutoff frequency. From this the separation between the parallel plates of the antenna 'a' was derived and set. The separation 'a' calculated for frequencies above 20 GHz is 7.5 mm, which is very much suitable for tapping in the frequencies of interest namely 28 GHz and 31 GHz. The same calculated parallel plate separation 'a' is retained for the other two outer ports, given, their desired cutoff frequencies are above the designed 20 GHz. This helps in attaining a shared aperture configuration for the three ports with multiple frequency operation.

*B. Design of the feeding horns*

The parallel plate waveguide with a derived separation of 7.5 mm between the plates acts as the driving factor in deciding the aperture dimensions of the horns that feed into the structure. An E-plane sectoral horn whose opening is flared in the direction of the E field, while keeping the other constant was chosen. The front aperture dimensions of the E-Plane sectoral horn were chosen as 15 mm x 7.5 mm and the back-wall dimensions were chosen as 5 mm x 7.5 mm with a flaring between the two faces provided over a length of 14 mm. The selection of these dimensions, that are dependent on the parallel plate waveguide dimension ensures co-design of the horn and the parallel plate waveguide interface; and the input matching of the sectoral horn at 28 GHz and 31 GHz. The feeding horns are further extended from their back walls by means of a propagating waveguide section at the designed frequency. The length of this waveguide section is chosen to have all the three port planes originating at the same plane. This provides for the ease and uniformity of feeding the ports, especially in an array.

The horns at the outer ends are spaced from the center horn so that a minimal spatial footprint is obtained while retaining the obtainable beamwidth. The outer horns are tilted towards the center to target the design angle of 60° and a waveguide section, suitably chamfered to reduce losses, connects these outer end horns to their feeding port plane.

*C. Design of the ridge based lens*

The wavefront of the $E_x$ component of the field emitted by the E-plane sectoral horn is cylindrical and this is converted to planar wavefront by means of a metal ridge-based lens. Phase correction is performed over the path lengths of the rays traced over the parallel plate waveguide, from the sectoral horn towards the lens. The variable that introduces the required delay for the correction is the height of the ridge. The initial distance from the E-plane sectoral horn aperture to the ridge is set using the azimuthal beamwidth of the E-plane horn at -12 dB and by calculating the corresponding arc length and radius ratio, using the trigonometric rule:

$$\theta = \frac{l}{r} \quad (5)$$

where, $\theta$ is the subtended beamwidth angle in radians and $l$ is the arc length subtended and $r$ is the radius. Using the above rule, the distance 'r' between the E-plane horn aperture and the ridge was set to 50 mm, which is in the Far-field region of the E-plane sectoral horn. The Far-field distance is 47.8 mm calculated for the horn aperture of 15 mm x 7.5 mm at 28 GHz. This further for a -12 dB beamwidth of 66° (1.15 radians) of the E-plane sectoral horn, necessitates a ridge length of 57 mm for optimal illumination.

The position and the length of the ridge being set, the height of the ridge can be calculated by considering the path length correction required to convert the cylindrical wavefront to planar wavefront at the interface of the ridge. The phase plot for the $E_x$ field component over the parallel plate waveguide floor is extracted using full wave solver and the maximum difference in phase between the phase of the emitted cylindrical wavefront at its center and that at the edge is calculated. This difference is then divided by the calculated phase gradient (constant over small distances) to attain the height of the ridge. It must be noted that intuitively, this suggests a varying ridge height along the section. However, here an average height derived from the maximum phase differences is considered. This has been performed to meet three essential requirements: to keep the ridge lens design simple, to offer similar input matching at the three ports and to have a stable radiation pattern over the three ports. The average height can not produce the same effects as the tapering of the ridge height but simplifies the fabrication process. The deviation seen is a pointing error and an increase of phase aberrations, that manifest as the asymmetric increase of sidelobes as seen in Fig. 6 and 7, but to a smaller extent.

### D. Design of the flared radiating aperture

The antenna designed through the above steps has a pointing fan beam in the elevation plane. The gain of the antenna can be improved by 1-2 dBi by providing a guiding section that flares out at an optimal flaring angle to further enhance the gain. The flaring is provided at the aperture of the waveguide section that follows the ridge lens. The change in the flaring angle causes a change in the path lengths between the waveguide and the aperture axially along the yz plane. The path difference is smallest for the case of flaring angle 0° measured outward along yz plane and it is maximum for the optimal flare angle. The flare angle cannot be increased beyond a limit as the phase distribution becomes predominantly quadratic and the TE10 fields do not add up constructively to enhance the gain and directivity. At the optimal flare angle there is a flat phase distribution at the aperture and the TE10 mode fields add up constructively to increase the gain. The optimal flaring angle in the design was chosen as 30° considering the phase distribution at the aperture. The flaring angle was set in such a way to have a flat phase distribution and reduce the quadratic phase distribution. It must be noted that directional variation in the way the aperture is fed has little or no effect on the directivity of the antenna, which reinforces the importance of the flaring angle as a parameter to enhance gain.

## IV. THREE PORT ANTENNA CHARACTERISTICS

The antenna was simulated in Ansys HFSS, a full wave EM solver. The antenna built on the principles depicted in section III was extended to create a three-port model. The material of the antenna was chosen as aluminium. It is designed for dual band operation in the frequency bands 28 GHz and 31 GHz with a bandwidth of 1 GHz each. The following sections presents the simulated return loss and the directional gain patterns of the antenna.

### A. Return loss

The three-port antenna has the S11, S22 and S33 < -10 dB around the 28 GHz and 31 GHz band. The band of operation is designed between 27.5 GHz to 28.5 GHz for the first band and between 30.5 GHz and 31.5 GHz for the second band at each of the three ports. The return losses due to the ports 2 and 3 have similar characteristics as compared to the central port 1 as shown in Fig. 4.

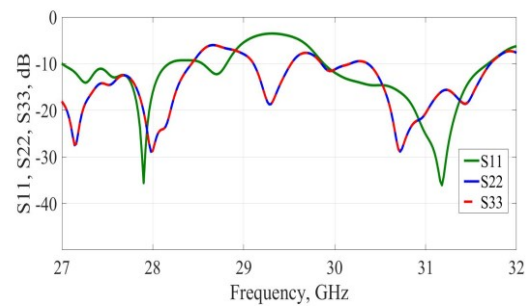

Fig. 4. Return loss at the three ports of the antenna

### B. Isolation between the ports

The proposed antenna is designed to act as a transceiver at two different independent frequencies. It can transmit at 28 GHz through the central port and receive at 31 GHz at the other two ports. To make this possible, the antenna has been designed with a port-to-port isolation less than -25 dB between the transmitter and the receiver. The transmitter and the receiver of the antenna have the same polarization; therefore, the isolation has been improved by arranging the two outer ports to face each other through their horns in a cross fashion. The port-to-port isolation graph is shown in Fig. 5.

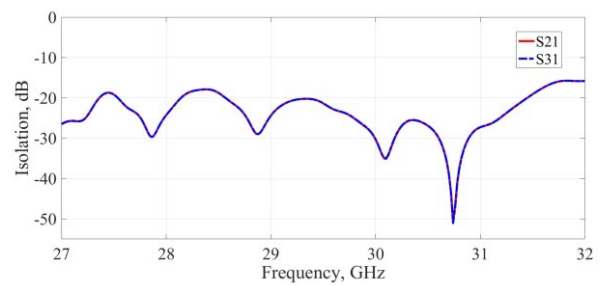

Fig. 5. Port-to-port isolation

## C. Radiation efficiency

In the design of millimeter wave antenna systems, the radiation efficiency is an important characteristic. In case of antenna acting as a transmitter it represents the portion of the input power delivered to free space by the transmitter antenna. At the receiver it represents the amount of the received power at the aperture of the antenna that is delivered to the receiver system. A good radiation efficiency implies longer range of transmission for a transmitter antenna. The designed antenna is intended to operate both as a transmitter and a receiver. The 3D EM solver calculated values of radiation efficiency at the two frequencies of 28 GHz and 31 GHz are shown in the Table 1 for all the three ports. The values depict a substantial value of the radiation efficiency.

| Frequency in GHz | Radiation efficiency in % | | |
|---|---|---|---|
| | Port 1 | Port 2 | Port 3 |
| 28 | 96.6 | 97.1 | 97.3 |
| 31 | 98.4 | 94.9 | 95.1 |

Table 1. Simulated radiation efficiency at the three ports over the dual band

The values are high since the structure is a waveguide operating in millimeter wave frequencies and does not contain substrates that introduce losses. In fact, this was a primary reason in choosing the waveguide-based lens antenna. Additional actual drop in radiation efficiency may be attributed to the losses introduced at the connectors and those due to fabrication tolerances.

## D. Directional gain pattern

The Multi-beam antenna casts beams along directions of +63°, 0° and -63°, which are within the tolerance limit of 5° for use within a sector cell with minimal sector-sector and inter-cell interference. Though the direction of the beams is so chosen as to cover the sector of a typical base station cell, they can be reconfigured to any desired angle. The gain obtained along the three beams for the two frequencies of operation are shown in the Table 2. The azimuthal beam pattern at 28 GHz and 31 GHz for the three ports along the azimuth is shown in Fig. 6 and 7. The elevation gain pattern at 28 GHz and 31 GHz is shown for the three ports in Fig. 8 and 9. The beam is fan shaped along the yz plane. It is to be noted that these beams are formed on the same shared aperture. The antenna at each of the three ports is horizontally polarized.

| Frequency in GHz | Beam gain in dBi | | |
|---|---|---|---|
| | Port 1 | Port 2 | Port 3 |
| 28 | 13.8 | 15.5 | 15.5 |
| 31 | 15.8 | 14.4 | 14.4 |

Table 2. Simulated beam gains at the three ports over the dual band

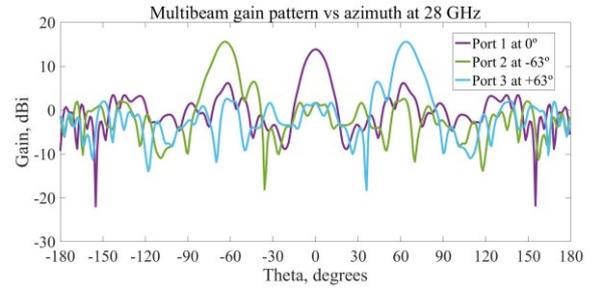

Fig. 6. Multibeam azimuthal gain pattern at 28 GHz for the three ports at Phi=0 plane

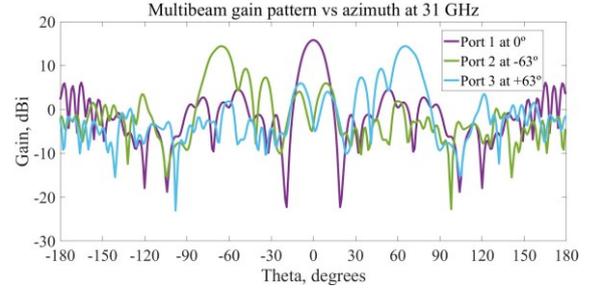

Fig. 7. Multibeam azimuthal gain pattern at 31 GHz for the three ports at Phi=0 plane

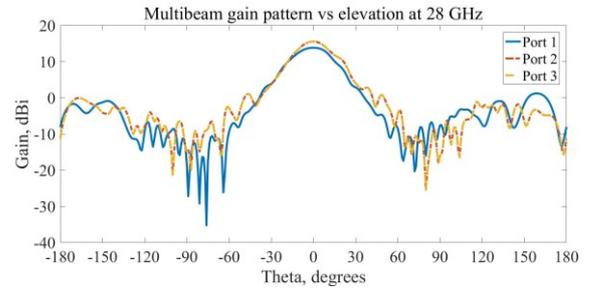

Fig. 8. Multibeam elevation gain pattern at 28 GHz for the three ports at Phi=90 plane

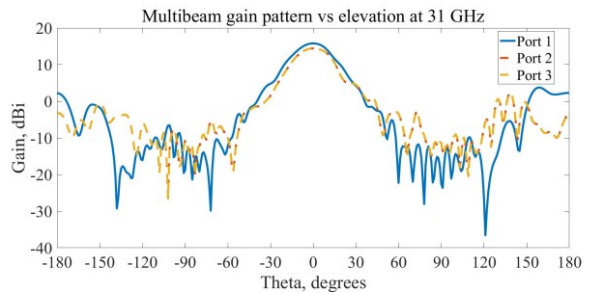

Fig. 9. Multibeam elevation gain pattern at 31 GHz for the three ports at Phi=90 plane

## V. Conclusions

The design and validation by full-wave simulations of a multi-beam dual band parallel plate beamformer has been presented. The antenna design has been shown to be a co-design process of four interdependent designs integrated to operate in unison. The antenna has an operational band in the 27.5 GHz to 28.5 GHz and the 30.5 to 31.5 GHz band. This dual band operation makes the antenna a potential candidate for use in the Local Multipoint Distribution System (LMDS) using millimeter wave for 5G. The three-port gain within the frequency band has been calculated by full wave simulations as being around 15 dBi. The antenna beams are directed at +60°, 0° and -60°, however using the proposed design methodology other beam directions can be obtained. The current design beam angles can be used in millimeter wave based 5G LMDS small cell scenarios which require an angular projection of 120° in each sector of the cell. Other potential use of the antenna is in the rotational surveillance radars, when used as a circular array of three 120º sectors. The general design methodology presented here allows to custom design this antenna for suitable high gain multi-beam applications. A prototype of this antenna is currently under manufacture. Manufacturing tolerances have been considered in the design and the proposed antenna will be validated in a full test campaign.

## Acknowledgment

This work was supported by the 5Gwireless project that has received funding from the European Union's Horizon 2020 research and innovation programme under grant agreement No. 641985

## References

[1] W. E. Kock, "Metallic delay lenses," in *The Bell System Technical Journal*, vol. 27, no. 1, pp. 58-82, Jan. 1948.

[2] J. Ruze, "Wide-Angle Metal-Plate Optics," in *Proceedings of the IRE*, vol. 38, no. 1, pp. 53-59, Jan. 1950.

[3] W. Rotman and R. Turner, "Wide-angle microwave lens for line source applications," in *IEEE Transactions on Antennas and Propagation*, vol. 11, no. 6, pp. 623-632, November 1963.

[4] Y. Zeng and R. Zhang, "Cost-Effective Millimeter-Wave Communications with Lens Antenna Array," in *IEEE Wireless Communications*, vol. 24, no. 4, pp. 81-87, 2017.

[5] P. Alitalo, F. Bongard, J. Mosig and S. Tretyakov, "Transmission-line lens antenna with embedded source," *2009 3rd European Conference on Antennas and Propagation*, Berlin, 2009, pp. 625-629.

[6] M. Ettorre, R. Sauleau and L. Le Coq, "Multi-Beam Multi-Layer Leaky-Wave SIW Pillbox Antenna for Millimeter-Wave Applications," in *IEEE Transactions on Antennas and Propagation*, vol. 59, no. 4, pp. 1093-1100, April 2011.

[7] *H. Legay* et al., "Multiple beam antenna based on a parallel plate waveguide continuous delay lens beamformer," *2016 International Symposium on Antennas and Propagation (ISAP),* Okinawa, 2016, pp. 118-119.

[8] F. Doucet, N. J. G. Fonseca, E. Girard, H. Legay and R. Sauleau, "Analysis and design of a continuous parallel plate waveguide multiple beam lens antenna at Ku-band," *2017 11th European Conference on Antennas and Propagation (EUCAP)*, Paris, 2017, pp. 3631-3635.

[9] F. Doucet, S. Tubau, E. Girard, N. Fonseca, H. Legay and R. Sauleau "Design of continuous parallel plate waveguide lens-like beamformers for future low-cost and high performances multiple beam antennas", *38th ESA Antenna Workshop on Innovative Antenna Systems and Technologies for Future Space Missions*, October 2017, ESTEC, Noordwijk, The Netherlands .

[10] D. M. Pozar, Microwave Engineering, 4th Edition, Wiley, 2011.